\documentclass[aps,prl,twocolumn,showpacs,floatfix]{revtex4} 
\usepackage{amsmath,amssymb,isolatin1,graphicx,times}

\newcommand{\be}{\begin{equation}}
\newcommand{\ee}{\end{equation}}
\newcommand{\bea}{\begin{eqnarray}}
\newcommand{\eea}{\end{eqnarray}}
\newcommand{\bw}{\begin{widetext}}
\newcommand{\ew}{\end{widetext}}
\newcommand{\kommentar}[1]{}
\newcommand{\pavg}{\Pi_M(t)}

\begin{document}
 
\title{Slow Excitation Trapping in Quantum Transport with Long-Range
Interactions}
\author{Oliver M{\"u}lken}
\email{muelken@physik.uni-freiburg.de}
\author{Volker Pernice}
\author{Alexander Blumen}
\affiliation{
Theoretische Polymerphysik, Universit\"at Freiburg,
Hermann-Herder-Stra{\ss}e 3, 79104 Freiburg, Germany}

 
\date{\today} 
\begin{abstract}
Long-range interactions slow down the excitation trapping in quantum
transport processes on a one-dimensional chain with traps at both ends.
This is counter intuitive and in contrast to the corresponding classical
processes with long-range interactions, which lead to faster excitation
trapping. We give a pertubation theoretical explanation of this effect.
\end{abstract}
\pacs{
05.60.Gg, 
05.60.Cd, 
71.35.-y 
}
\maketitle


Building a quantum system from scratch has become possible due to recent
experimental advances in controlling and manipulating atoms and molecules.
It has actually become possible to tailor theoreticians favourite
one-dimensional systems using, e.g., ultra-cold atoms in optical lattices,
see \cite{bloch2005} and references therein.  From a dynamical point of
view, this allows for these systems to compare the theoretical predicitons
for the transport of charge, mass, or energy to the experimental results.
In turn, the experimental findings might eventually lead to a refinement
of the theoretical models. 

The tight-binding approximation for the transport of a quantum particle
over a regular structure (network) is a simple description which is
equivalent to the so-called continous-time quantum walks (CTQW) with
nearest-neighbor interactions (NNI) \cite{farhi1998,mb2005a}.  Recently,
several experiments have been proposed addressing CTQW, e.g., based on
wave guide arrays \cite{hagai2007}, atoms in optical lattices
\cite{duer2002,cote2006}, or structured clouds of ultra-cold Rydberg atoms
\cite{mbagrw2007}. In some of these experiments one finds long-range
interactions (LRI), such as in Rydberg gases, where also blockade
\cite{lukin2001} and antiblockade \cite{ates2007a} effects have to be
considered.  In a recent study of the effect of LRI on the quantum
dynamics in a linear system it has been found that CTQW for all
interactions decaying as $R^{-\nu}$ (where $R$ is the distance between two
nodes of the network) belong to the same universality class for $\nu>2$,
while for classical continuous-time random walks (CTRW) universality only
holds for $\nu>3$ \cite{mpb2008a}.

Coupling a system to an absorbing site, i.e., to a trap, allows to monitor
the transport by observing the decay of the survival probability of the
moving entity, say, the excitation. In the long-time limit and for NNI the
decay is practically exponential for both, classical systems modeled by
CTRW \cite{klafter1980} and quantum systems modeled by CTQW
\cite{excitontrap,mbagrw2007}. At intermediate times, which are
experimentally relevant, there appear considerable, characteristic
differences between the classical and the quantum situations
\cite{mbagrw2007}. 


Here, we study the quantum dynamics of one-dimensional CTQW with LRI in
the presence of traps and use the similarity to CTRW for a comparison to
the respective classical case.  Without traps, we model the quantum
dynamics on a network of connected nodes by a tight binding Hamiltonian
${\bf H}_0$.  For the corresponding classical process, we identify the
CTRW transfer matrix ${\bf T}_0$ with ${\bf H}_0$, i.e., ${\bf H}_0 = -
{\bf T}_0$; see e.g.\ \cite{farhi1998,mb2005a} for details.  For
undirected networks, ${\bf H}_0$ is related to the connectivity matrix
${\bf A}_0$ of the network by ${\bf H}_0 = {\bf A}_0$.  When the
interactions between two nodes go as $R^{-\nu}$, with $R=|k-j|\geq1$ being
the distance between two nodes $j$ and $k$, the Hamiltonian has the
following structure:
\bea 
{\bf H}_0(\nu) &=& \sum_{n=1}^N \Bigg[ \sum_{R=1}^{n-1} R^{-\nu} \Big( | n
\rangle \langle n| - | n-R \rangle \langle n | \Big) \nonumber \\ && +
\sum_{R=1}^{N-n} R^{-\nu} \Big( | n \rangle \langle n| - | n+R \rangle
\langle n | \Big) \Bigg].
\label{hamil_long}
\eea
We restrict ourselves to extensive cases ($\nu>1$), i.e., we explicitly
exclude ultra-long range interactions.  The corresponding NNI Hamiltonian
is obtained for $\nu=\infty$, in which case only the leading terms with
$R=1$ do not vanish.

The states $|j\rangle$ associated with excitations localized at the nodes
$j$ ($j=1,\dots,N$) form a complete, orthonormal basis set of the whole
accessible Hilbert space ($\langle k | j \rangle = \delta_{kj}$ and
$\sum_k |k~\rangle\langle~k| = {\bf 1}$). In general, the transition
probabilities from a state $|j\rangle$ at time $t_0=0$ to a state
$|k\rangle$ at time $t$ read $\pi_{kj}(t) \equiv
\left|\alpha_{kj}(t)\right|^2 \equiv \left|\langle k | \exp[-i {\bf
H}_0(\nu) t] | j \rangle\right|^2$. In the corresponding classical CTRW
case the transition probabilities follow from a master equation as
$p_{kj}(t) = \langle k | \exp({\bf T}_0 t) | j \rangle$
\cite{farhi1998,mb2005a}.


Now, let the nodes $m$ ($m\in{\cal M}$ and ${\cal M}\subset\{1,\dots,N\}$)
be traps for the excitation.  Within a phenomenological approach, the new
Hamiltonian is ${\bf H}(\nu) \equiv {\bf H}_0(\nu) - i{\bf \Gamma}$, with
the trapping operator $i{\bf \Gamma} \equiv i \Gamma \sum_{m\in{\cal M}} |
m \rangle \langle m |$, see Ref.~\cite{mbagrw2007} for details.  As a
result, ${\bf H}$ is non-hermitian and has $N$ complex eigenvalues, $E_l =
\epsilon_l - i\gamma_l$ ($l=1,\dots,N$) with $\gamma_l>0$, and $N$ left
and $N$ right eigenstates, denoted by $|\Psi_l\rangle$ and
$\langle\tilde\Psi_l|$, respectively.  The transition probabilities follow
as 
\be
\pi_{kj}(t) = \Big|\sum_l \exp(-\gamma_lt) \exp(-i\epsilon_lt)\langle k |
\Psi_l \rangle \langle \tilde\Psi_l | j \rangle\Big|^2,
\ee
where the imaginary parts $\gamma_l$ of $E_l$ determine the temporal
decay. For the incoherent classical process the description by CTRW is
quite similar: The new transfer operator reads ${\bf T}(\nu) = {\bf
T}_0(\nu) - {\bf \Gamma} = - {\bf A}_0(\nu) - {\bf \Gamma}$, which is real
and symmetric, leading to the eigenvalues $-\lambda_l$ ($\lambda_l>0$) and
corresponding eigenstates $|\Phi_l\rangle$. Note that due to the different
incorporation of the trapping operator in ${\bf T}(\nu)$ and ${\bf
H}(\nu)$ the corresponding eigenvalues and eigenstates will differ.
Without trapping we have ${\bf T}_0(\nu) = - {\bf H}_0(\nu)$ and thus
$\lambda_l \equiv E_l$ and $| \Phi_l \rangle \equiv | \Psi_l \rangle$.

In order to make a global statement for the whole network, we calculate
the mean survival probability for a total number of $M$ trap nodes, 
\be
\pavg \equiv \frac{1}{N-M} \sum_{j\not\in{\cal M}} \sum_{k\not\in{\cal M}}
\pi_{kj}(t), 
\ee
i.e., the average of $\pi_{kj}(t)$ over all initial nodes $j$ and all
final nodes $k$, neither of them being a trap node.  Classically, we will
consider $P_M(t) \equiv 1/(N-M) \sum_{j\not\in{\cal M}}
\sum_{k\not\in{\cal M}} p_{kj}(t)$.  For intermediate and long times and a
small number of trap nodes, $\pavg$ is mainly a sum of exponentially
decaying terms \cite{mbagrw2007}:
\be
\pavg \approx \frac{1}{N-M} \sum_{l=1}^N \exp(-2\gamma_lt).
\label{pi_avg2}
\ee
If the imaginary parts $\gamma_l$ obey a power-law with an exponent $\mu$
($\gamma_l \sim a l^\mu$), the mean survival probability scales as $\pavg
\sim t^{-1/\mu}$.

\begin{figure}[htb]
\centerline{
\includegraphics[clip=,width=0.9\columnwidth]{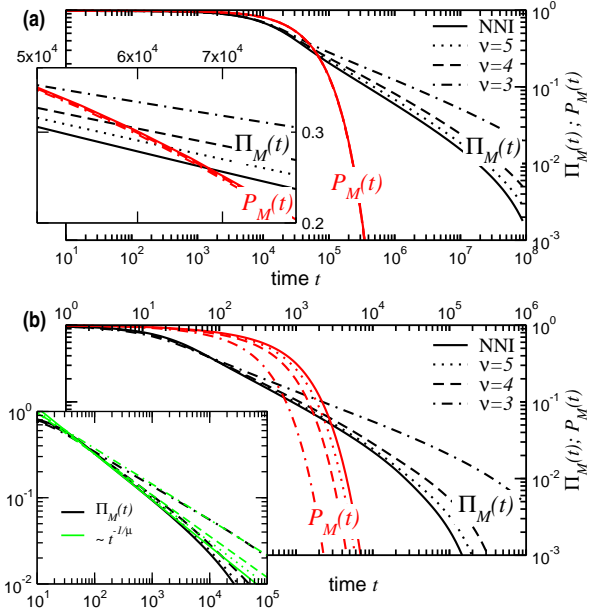}
}
\caption{(Color online) $\nu$-dependence of the quantum mechanical $\pavg$
and the classical $P_M(t)$ decay behaviors for a chain of $N=100$ sites;
here (a) $\Gamma=0.001$ and (b) $\Gamma=1$. The inset in (a) shows a
close-up picture of the region where $\pavg$ and $P_M(t)$ cross. The inset in (b)
shows power-law fits to $\pavg$ in the intermediate time regime with
exponents $1/\mu$, where the $\mu$ are taken from
Fig.~\ref{evals_imag_longrange}(b).}
\label{decay_n100}
\end{figure}

In Ref.~\cite{mbagrw2007} an experimental setup was proposed, which is
based on a finite linear chain of clouds of ultracold Rydberg atoms with
trapping states at both ends ($m=1,N$). There, the dynamics was
approximated by a NNI tight binding model, which - for a ring without
traps - has been shown to behave in the same fashion as systems with LRI
of the form $R^{-\nu}$ for which $\nu>2$ \cite{mpb2008a}.  The Rydberg
atoms interact via dipole-dipole forces, i.e., the potential between two
atoms decays roughly as $R^{-3}$.

For the finite chain with $m=1,N$, Fig.~\ref{decay_n100} shows a
comparison of the quantum mechanical $\pavg$ and of the classical $P_M(t)$
behaviors for different $\nu$ and $\Gamma$, which were obtained by
numerically diagonalizing the corresponding Hamiltonian ${\bf H}(\nu)$ and
transfer matrix ${\bf T}(\nu)$, respectively. Clearly, for both
$\Gamma$-values the LRI lead to a slower decay of $\pavg$, i.e., to a
slower trapping of the excitation, which is counter intuitive since the
opposite effect is observable for classical systems where the decay of
$P_M(t)$ becomes faster for decreasing $\nu$, see below. By increasing the
trapping strength $\Gamma$, the difference between the quantum and the
classical behavior becomes even more pronounced, compare
Figs.~\ref{decay_n100}(a) and \ref{decay_n100}(b).  Generally, for $\pavg$
the change in $\Gamma$ results mainly in a rescaled time axis, since the
imaginary parts $\gamma_l$ are of the same order of magnitude when
rescaled by $\Gamma$. For the specific case of the Rydberg atoms ($\nu=3$
and $\Gamma=1$) one observes the largest difference between the $\pavg$
and the $P_M(t)$ behaviors.  To understand this phenomenon, we continue to
analyze $\pavg$ within a perturbation theoretical treatment. 


When the strength of the trap, $\Gamma$, is small compared to the
couplings between neighboring nodes, we can evaluate the eigenvalues using
perturbation theory, see, for instance, \cite{Sakurai}. Let
$|\Psi^{(0)}_l\rangle$ be the $l$th eigenstate and $E^{(0)}_l\in
\mathbb{R}$ be the $l$th eigenvalue of the unperturbed system with
Hamiltonian ${\bf H}_0(\nu)$. Up to first-order the eigenvalues of the
perturbed system are given by
\be
E_l = E^{(0)}_l - i \Gamma \sum_{m\in{\cal M}} \Big| \langle m
|\Psi^{(0)}_l\rangle \Big|^2.
\label{evals_perturb}
\ee
Therefore, the correction term determines the imaginary parts $\gamma_l$,
while the unperturbed eigenvalues are the real parts $\epsilon_l =
E^{(0)}_l$.  Having only a few trap nodes, the sum in
Eq.~(\ref{evals_perturb}) contains only few terms.  Moreover, from
Eq.~(\ref{evals_perturb}) we also see that the imaginary parts $\gamma_l$
are essentially determined by the eigenstates $|\Psi^{(0)}_l\rangle$ of
the system without traps. A change in these states will also lead to a
change in the $\gamma_l$. As we proceed to show, this is exactly what
happens by going from NNI to LRI.



Without loss of generality, an eigenstate of a finite chain with NNI can
be written as ($l=1,\dots,N$)
\be
|\Psi^{(0)}_l\rangle = 
\begin{cases}
\displaystyle
\sqrt{\frac{1}{N}} \sum_{j=1}^N | j\rangle & l=N  \\
\displaystyle
\sqrt{\frac{2}{N}} \sum_{j=1}^N \cos\big[(2j-1)\theta_l/2\big] |j\rangle &
\mbox{else},
\end{cases}
\label{statesNN}
\ee
where for convenience we take $\theta_l \equiv \pi(N-l)/N \in [0,\pi[$;
the corresponding eigenvalues are $E^{(0)}_l = 2 - 2\cos\theta_l$ (note
that the smallest eigenvalue is $E^{(0)}_N=0$). Thus, to first order
perturbation theory we obtain from Eqs.~(\ref{evals_perturb}) and
(\ref{statesNN}) as imaginary parts $\gamma_N=2\Gamma/N$ and $\gamma_l =
(4\Gamma/N) \cos^2\big(\theta_l/2\big) = (2\Gamma/N) [1+\cos\theta_l]$ for
$l=1,\dots,N-1$, which for $l \ll N$ yields $\gamma_l \sim l^2$. In this
case the mean survival probability will scale in the corresponding time
interval as $\pavg \sim t^{-1/2}$. 

Formally, we can perform the continuum limit $N\to\infty$ (by taking now
$4\Gamma/N \equiv a$ finite). Then the sum in Eq.~(\ref{pi_avg2}) turns
into an integral such that 
\be
\pavg \sim e^{-at} \ \frac{1}{\pi}\int\limits_0^\pi d\theta \ e^{-at
\cos\theta} =e^{-at} I_0(at),
\ee
where $I_0(at)$ is the modified Bessel function of the first kind
\cite{abramowitz}. From this we get for large $t$ that $\pavg\sim
t^{-1/2}$, which confirms the previous results.  Note, however, that for
small $N$ the smallest $\gamma_l$-value is finite and, therefore, the
scaling of $\gamma_l$ holds only in a quite small interval of $l$-values.
Hence, also the time interval in which $\pavg$ scales with the exponent
$-1/2$ is rather small. A lower bound for scaling is given by the behavior
of $\gamma_l$ for $l\approx N/2$ (corresponding to smaller times than for
$l \ll N$). Here, $\gamma_l$ is linear in $l$, which leads to a lower
bound of $\mu\geq 1$ for the scaling exponent.  An exponent $\mu$ which is
valid over a larger $l$-interval will therefore be in the interval $[1,2]$
and, consequently, the exponent for $\pavg$ will lie in the interval
$[-1,-1/2]$. 


In the case of periodic boundary conditions, one finds
translation-invariant Bloch eigenstates regardless of the range of the
interaction \cite{mpb2008a}.  In the case of Eq.~(\ref{hamil_long}),
however, the eigenstates for LRI differ from the ones for NNI
[Eq.~(\ref{statesNN})]; In Eq.~(\ref{hamil_long}) the finite extension of
the chain destroys the translational invariance. As is immediately clear
from Eq.~(\ref{evals_perturb}), this also implies that the imaginary parts
of the eigenvalues, evaluated based on first order perturbation theory,
will change.

For large exponents $\nu$ we can regard the LRI as a small perturbation to
the NNI, i.e., having ${\bf H}_0(\nu) = {\bf H}_0 + {\bf H}_{\nu}$, where
${\bf H}_\nu$ contains only the correction terms to the NNI case ${\bf
H}_0$. This allows us to calculate from the unperturbed states
$|\Psi^{(0)}_l\rangle$ the perturbed eigenstates $|\Psi_l\rangle$ up to
first order. Taking the states $|\Psi_l\rangle$ to be the eigenstates of
the LRI system without traps, we readily obtain the imaginary parts
$\gamma_l$ for small trapping strength from Eq.~(\ref{evals_perturb}) as
$\gamma_l = 2\Gamma  \big| \langle 1 |\Psi_l\rangle \big|^2$, where
\be
\langle 1 | \Psi_l\rangle = \langle 1 | \Psi^{(0)}_l\rangle + \sum_{r\neq
l} \frac{ \langle \Psi^{(0)}_r | {\bf H}_\nu | \Psi^{(0)}_l
\rangle}{E^{(0)}_l - E^{(0)}_r} \langle 1 | \Psi^{(0)}_r\rangle.
\label{states_lr}
\ee 

It is straightforward, although cumbersome, to calculate the corrections
to the imaginary parts $\gamma_l$ from Eq.~(\ref{states_lr}).  For large
$\nu$ the coupling to the next-next-nearest neighbor is by a factor of
$(3/2)^\nu$ smaller, for $\nu=10$ this is about one and a half orders of
magnitude. Taking, for fixed $\nu$, only nearest and next-nearest neighbor
couplings into account allows us to obtain simple analytic expressions.
The perturbation term ${\bf H}_{\nu}$ is now tri-diagonal.  Its non-zero
elements are $\langle j-2 | {\bf H}_\nu | j \rangle = \langle j+2 | {\bf
H}_\nu | j \rangle = - 2^{-\nu}$ and its diagonal elements follow from
$\langle j | {\bf H}_\nu | j \rangle = - \sum_i \langle i | {\bf H}_\nu |
j \rangle$, thus $\langle j | {\bf H}_\nu | j \rangle = 2^{-\nu}$ for
$2<j<N-1$ and $\langle j | {\bf H}_\nu | j \rangle = 2^{-\nu+1}$ else. We
hence obtain from Eq.~(\ref{states_lr})
\be
\langle 1 | \Psi_l\rangle =
\sqrt{\frac{2}{N}}\cos\Big(\frac{\theta_l}{2}\Big) + 2^{-\nu}
\sqrt{\frac{2}{N}} \sin\big(2\theta_l\big)
\sin\Big(\frac{\theta_l}{2}\Big).
\label{states_lr_approx}
\ee 

\begin{figure}[htb]
\centerline{
\includegraphics[clip=,width=0.9\columnwidth]{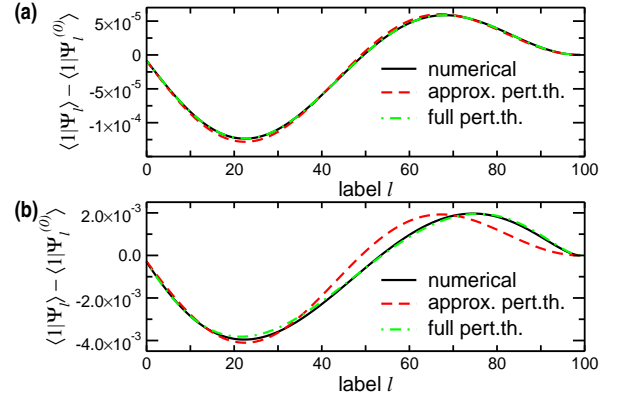}
}
\caption{(Color online) Correction term $\langle 1 | \Psi_l\rangle -
\langle 1 | \Psi^{(0)}_l\rangle$ for $N=100$ and for (a) $\nu=10$ and (b)
$\nu=5$. The direct numerical evaluation (solid black line) is compared
to the perturbation theory expression Eq.~(\ref{states_lr})
(dashed-dotted green line) and to the approximate expression
Eq.~(\ref{states_lr_approx}) (dashed red line).
}
\label{pertub_nu}
\end{figure}

Figure~\ref{pertub_nu} shows the difference $\langle 1 | \Psi_l\rangle -
\langle 1 | \Psi^{(0)}_l\rangle$ for $N=100$ and for (a) $\nu=10$ and (b)
$\nu=5$. The numerical exact value (solid black line) is obtained by
computing separately $\langle 1 | \Psi_l\rangle$ and $\langle 1 |
\Psi^{(0)}_l\rangle$ and subsequently taking the difference; the result is
then confronted to Eq.~(\ref{states_lr}) (dashed-dotted green line),
determined numerically, and to Eq.~(\ref{states_lr_approx}) (dashed red
line).  For $\nu=10$, the agreement between all three curves is remarkably
good, see Fig.~\ref{pertub_nu}(a), which justifies the assumptions leading
to Eq.~(\ref{states_lr_approx}). For smaller $\nu$ [$\nu=5$ in
Fig.~\ref{pertub_nu}(b)] there is still a reasonable agreement between
Eq.~(\ref{states_lr}) and the exact result; however, taking only nearest
and next-nearest neighbors into account leads to evident deviations, see
the dashed red line in Fig.~\ref{pertub_nu}(b). 

Now, from Eq.~(\ref{states_lr_approx}) we get 
\be
\gamma_l \approx \gamma_l^{(0)} + 2^{-\nu} \gamma_l^{(1)} + {\cal
O}(2^{-2\nu}),
\label{gamma_approx}
\ee
where $\gamma_l^{(0)}$ is the NNI expression given above and
$\gamma_l^{(1)} = (8\Gamma/N) \cos\big(\theta_l/2\big)
\sin\big(2\theta_l\big) \sin\big(\theta_l/2\big)$ the correction due to
the LRI.  Again, the smallest $\gamma_l$-values are those for which $l \ll
N$, which leads to a decrease of the imaginary parts $\gamma_l$ because
$\gamma_l^{(1)} < 0$ for $l \ll N$.  Here, one can approximate the
imaginary parts by a power-law, i.e., $\gamma_{ l} \sim l^\mu$. A rough
estimate of the scaling exponent $\mu$, assuming $\nu\gg1$ can be readily
given. For this we note that from Eq.~(\ref{gamma_approx}) we have
$\ln\gamma_{l+1} - \ln\gamma_{l} \approx \ln\gamma_{l+1}^{(0)} -
\ln\gamma_{l}^{(0)} + 2^{-\nu} \big[(\gamma_{l+1}^{(1)}/\gamma_{l+1}^{(0)}
- \gamma_{l}^{(1)}/\gamma_{l}^{(0)}\big]$.  Moreover, the term $\mu^{(0)}
\equiv \big[\ln\gamma_{l+1}^{(0)} - \ln\gamma_{l}^{(0)}\big] / \big[\ln
(l+1) - \ln l\big]$ gives the exponent for the NNI case and the term
$\mu^{(1)} \equiv \big[\gamma_{l+1}^{(1)}/\gamma_{l+1}^{(0)} - \gamma_{
l}^{(1)}/\gamma_{l}^{(0)}\big]/\big[\ln (l+1) - \ln l\big]$ is the LRI
correction. Thus
\be
\mu \approx
\frac{\ln\gamma_{l+1} - \ln\gamma_{l}}{\ln
(l+1) - \ln l} \approx \mu^{(0)} + 2^{-\nu} \mu^{(1)}
\ee
Since $\mu^{(1)}$ is strictly positive for small $l$, the inclusion of LRI
leads to a decrease of $\gamma_{l}$ when compared to the NNI case. In
turn, this results in a slower decay of $\pavg$.

\begin{figure}[htb]
\centerline{
\includegraphics[clip=,width=0.9\columnwidth]{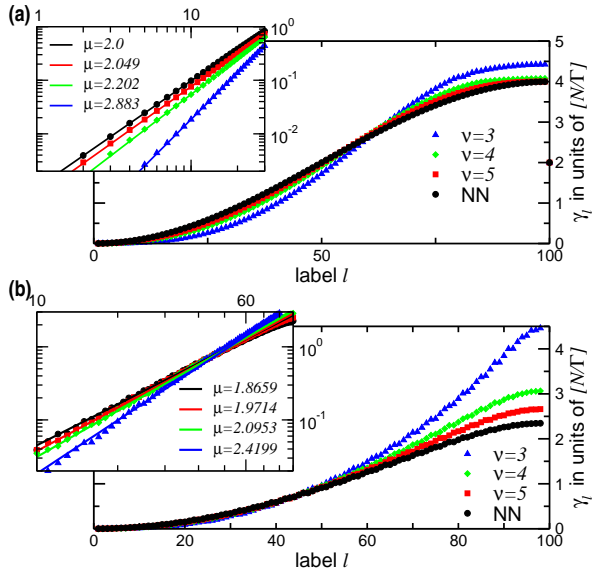}
}
\caption{(Color online) Imaginary parts $\gamma_l$ (dots) in ascending
order for LRI systems with $\nu=2$, $3$, $4$, and for NNI for $N=100$ and
(a) $\Gamma=0.001$ and (b) $\Gamma=1$.
}
\label{evals_imag_longrange}
\end{figure}

Figure~\ref{evals_imag_longrange} shows the imaginary parts $\gamma_l$ for
a chain of $N=100$ nodes with LRI ($\nu=3$, $4$, $5$) and with NNI. For
small $l$ and NNI, the $\gamma_l$ obey scaling with the exponent $\mu=2$,
as discussed above.  Introducting LRI, i.e., decreasing $\nu$, increases
the scaling exponent to $\mu>2$.  Consequently, the scaling exponent
$1/\mu$ for $\pavg$ decreases, leading to a slowing-down of the excitation
trapping due to LRI.  

In the classical case decreasing $\nu$ leads to a faster excitation
trapping, which is obervable in a quicker decay of $P_M(t)$.  This can
also be deduced from a perturbation theoretical treatment. As can be seen
from Fig.~\ref{decay_n100} (see also Fig.~2 of Ref.~\cite{mbagrw2007}),
the decay of $P_M(t)$ is exponential already at intermediate times and is
dominated by the smallest eigenvalue $\lambda_N$ and the corresponding
eigenstate $|\Phi_N\rangle$ of the transfer operator ${\bf T}(\nu)$: 
\bea
P_M(t) &=& \frac{1}{N-M} \sum_{l=1}^{N} \exp(-\lambda_l t) \Big|
\sum_{k\not\in{\cal M}} \langle k | \Phi_l \rangle \Big|^2 \nonumber \\
&\approx& \frac{1}{N-M} \exp(-\lambda_N t) \Big|
\sum_{k\not\in{\cal M}} \langle k | \Phi_N \rangle \Big|^2 .
\eea 
Calculating $\lambda_N$ and the prefactor $\big| \sum_{k\not\in{\cal M}}
\langle k | \Phi_N \rangle \big|^2$ for large $\nu$ and small $\Gamma$
shows that with decreasing $\nu$ the smallest eigenvalue $\lambda_N$
increases while the prefactor decreases. Together, this confirms our
numerical result of a quicker decay for $P_M(t)$, see
Fig.~\ref{decay_n100}.

Finally, we comment on the impact of our results on the experiment
proposed in Ref.~\cite{mbagrw2007}. Here, clouds of laser-cooled ground
state atoms are assembled in a chain by optical dipole
traps~\cite{grimm00}, which are then excited into a Rydberg S-state, see
\cite{mbagrw2007} for details.  The Rydberg atoms interact via long-range
dipole-dipole forces which is advantageous in many ways. As can be deduced
from Fig.~\ref{decay_n100}, the time intervals over which the decay
follows the power-law are enlarged by the LRI. For $\nu=3$ the transition
to the long-time exponential decay occurs at times which are about two
order of magnitude larger than the ones found for the NNI case. The
difference between a purely coherent (CTQW) and a purely incoherent (CTRW)
process is enlarged due to the LRI, allowing for a better discrimination
between the two when clarifying the nature of the energy transfer dynamics
in ultra-cold Rydberg gases.


In conclusion, we have considered the quantum dynamics of excitations with
LRI on a network in the presence of absorbing sites (traps). The LRI lead
to a slowing-down of the decay of the average survival probability, which
is counter intuitive since for the corresponding classical process one
observes a speed-up of the decay. Using pertubation theory arguments we
were able to identify the reason for this slowing-down; it results from
changes in the imaginary parts of the spectrum of the Hamiltonian.  

Support from the Deutsche For\-schungs\-ge\-mein\-schaft (DFG) and the
Fonds der Chemischen Industrie is gratefully acknowledged.

\end{document}